\newcommand{\ket}[1]{\mbox{\ensuremath{|#1\rangle}}}

\newcommand{\Kd}{\mbox{\ensuremath{\hat{K}^{\dagger}}}}
\newcommand{\K}{\mbox{\ensuremath{\hat{K}}}}
\newcommand{\Kdo}{\mbox{\ensuremath{\hat{K}_1^{\dagger}}}}
\newcommand{\Kdt}{\mbox{\ensuremath{\hat{K}_2^{\dagger}}}}
\newcommand{\ahd}{\mbox{\ensuremath{\hat{a}_h^{\dagger}}}}
\newcommand{\bhd}{\mbox{\ensuremath{\hat{b}_h^{\dagger}}}}
\newcommand{\avd}{\mbox{\ensuremath{\hat{a}_v^{\dagger}}}}
\newcommand{\bvd}{\mbox{\ensuremath{\hat{b}_v^{\dagger}}}}
\newcommand{\ah}{\mbox{\ensuremath{\hat{a}_h}}}
\newcommand{\bh}{\mbox{\ensuremath{\hat{b}_h}}}
\newcommand{\av}{\mbox{\ensuremath{\hat{a}_v}}}
\newcommand{\bv}{\mbox{\ensuremath{\hat{b}_v}}}

\documentclass[prl,twocolumn]{revtex4}

\usepackage{graphicx}

\begin{document}

\author{Antia Lamas-Linares}
\email{a.lamas@qubit.org}
\author{John C. Howell}
\email{john.howell@qubit.org}
\author{Dik Bouwmeester}
\email{dik.bouwmeester@qubit.org}

\affiliation{Centre for Quantum Computation, Clarendon Laboratory,
University of Oxford, Parks Road, Oxford OX1 3PU, UK}

\title{Stimulated emission of polarization--entangled photons}

\begin{abstract}
Entangled photon pairs -- discrete light quanta that exhibit
non-classical correlations -- play a crucial role in quantum
information science (for example in demonstrations of quantum
non-locality
\cite{clauser:78,aspect:82b,ou:88,shih:88,tittel:98,weihs:98,pan:00},
quantum teleportation \cite{bouwmeester:97,boschi:98} and quantum
cryptography \cite{ekert:92,jennewein:00,tittel:00,naik:00}). At
the macroscopic optical field level non-classical correlations can
also be important, as in the case of squeezed light
\cite{slusher:85}, entangled light beams \cite{wu:86,ou:92}, and
teleportation of continuous quantum variables \cite{furusawa:98}.
Here we use stimulated parametric down-conversion to study
entangled states of light that bridge the gap between discrete and
macroscopic optical quantum correlations. We demonstrate
experimentally the onset of laser-like action for entangled
photons. This entanglement structure holds great promise in
quantum information science where there is a strong demand for
entangled states of increasing complexity.
\end{abstract}

\maketitle

 As the acronym LASER (Light Amplification by Stimulated
Emission of Radiation) indicates, polarization--entangled laser
operation would mean that a (spontaneously created) photon pair in
two polarization--entangled modes stimulate, inside a non-linear
gain medium, the emission of additional pairs. As a gain medium we
consider type--II parametric down-conversion \cite{kwiat:95}. A
simplified interaction Hamiltonian \cite{kok:00, simon:00} for the
nonlinear interaction between a classical pump field and two
polarization--entangled modes $a$ and $b$ is given by
\begin{equation}
\hat{H}_{int}=e^{i\phi}\kappa \Kd+e^{-i\phi}\kappa \K
\label{eq:hamiltonian}
\end{equation}
where $\Kd=\ahd\bvd-\avd\bhd$ and $\K=\ah\bv-\av\bh$ are the
creation and annihilation operators of polarization entangled
photon pairs in modes $a$ and $b$. Horizontal and vertical
polarization are represented by H and V, and $\kappa$ is a
real-valued coupling coefficient. When acting on the vacuum state
the time evolution operator $\hat{U}=\exp{i\hat{H}t/\hbar}$
yields:
\begin{equation}
\ket{\psi}\propto\sum_{n=0}^{\infty}(\tanh \tau)^n
\sum_{m=0}^{n}(-1)^m\ket{n-m,m;m,n-m}\label{eq:state}
\end{equation}
where $\tau=\kappa t/\hbar$ is the interaction parameter. The
first and second slots in the ket indicate respectively the number
of horizontal $(n-m)$ and vertical $(m)$ photons in mode $a$,  and
the third and fourth slot indicate the corresponding numbers for
mode $b$. This state represents the general output of type-II
parametric down-conversion, but for all experiments reported to
date, $\tau$ is so small that mainly the first order term $(n=1)$
has been taken into account and only a few experiments and
proposals addressed second order terms
\cite{pan:00,ou:99,bouwmeester:99,demartini:00,demartini:01,weinfurter:01}.
By analogy with a conventional laser, the idea of an entangled
photon laser is to increase $\tau$  using a resonator around the
gain medium, which enhances the emission of the higher order terms
in equation \ref{eq:state}. The state shown in equation
\ref{eq:state} has the following features.

\begin{figure}
\scalebox{0.9}{\includegraphics{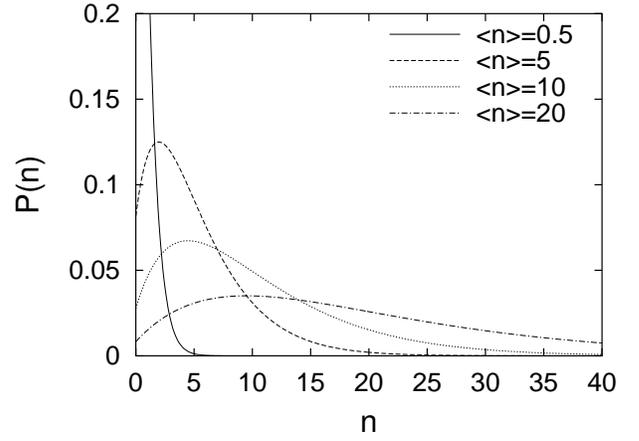}}
    \caption{The photon number (pair) distribution, $P(n)$,  arising
     from stimulated parametric down-conversion shifts its peak
      and broadens as the mean number of photons increases. This
       indicates that for increasing interaction strength (gain),
        terms with higher numbers of photons obtain a larger
        amplification factor compared to lower terms, which is
         a familiar feature of laser operation.} \label{fig:distribution}
\end{figure}

First, modes $a$ and $b$ are entangled in photon number since for
any $n$, the number of photons in each mode is identical. The
photon (pair) number distribution is shown in Fig.
\ref{fig:distribution} for increasing average photon (pair) number
output per pulse $\langle n \rangle$. The shifting of the maximum
and the broadening of the distribution for higher values of
$\langle n \rangle$ resembles the coherent state photon--number
distribution as produced by conventional lasers. These features
are explained by the fact that stimulated emission --originating
from the boson statistics of photons-- favours amplification of
higher over lower photon--number terms.

The second important property of state in equation \ref{eq:state}
is that the set of terms for each $n$ form a maximally entangled
state in polarization. The normalized 1--pair term is the
rotationally symmetric Bell state (singlet spin--$\frac{1}{2}$):
  \begin{equation}
    \ket{\psi}=\frac{1}{\sqrt{2}}(\ket{1,0;0,1}-\ket{0,1;1,0})
    \label{eq:singlet}
  \end{equation}
The normalized 2-pair term is given by :
 \begin{equation}
    \ket{\psi}=\frac{1}{\sqrt{3}}(\ket{2,0;0,2}-\ket{1,1;1,1}+\ket{0,2;2,0})
    \label{eq:spin1singlet}
  \end{equation}
and represents the singlet spin--1 state. Similar to the
spin--$\frac{1}{2}$) case, the rotational symmetry arises from the
relative phase relations and the equal weights of the terms. In
general the $n$--pair term has the properties of a singlet
spin--$n/2$ state. The rotational symmetry of the full state can
easily be shown by expressing the $n$--pair terms in any other
basis, and verifying that the same expressions are obtained. The
crucial role of stimulated emission is to provide for each $n$
equally weighted terms. In principle, a photon counting
measurement on state \ref{eq:state} (either in mode $a$ or $b$)
performs a projection onto a certain singlet spin--$n/2$ state.
Subsequently this maximally entangled state can be explored for
quantum information tasks. In practice, in quantum optics
experiments where the fragile photons are in general destroyed by
any measurement, the projection and the exploration of the state
are performed simultaneously. This procedure, usually referred to
as post-selection, has proven to be very useful: for example for
demonstrations of quantum teleportation \cite{bouwmeester:97},
quantum cryptography \cite{ekert:92, jennewein:00, tittel:00,
naik:00}, and three particle Greenberger--Horne--Zeilinger (GHZ)
correlations \cite{pan:00,bouwmeester:99}, and for novel optical
quantum computation schemes \cite{knill:01}. Here we use
post--selection to demonstrate stimulated entanglement by
measuring 2-- and 4--photon properties of state \ref{eq:state} for
increasing values of $\tau$.

\begin{figure}
\scalebox{.5}{\includegraphics{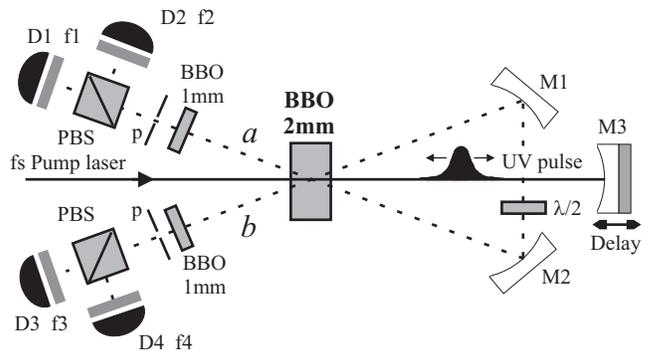}} \caption{Experimental
set--up. A frequency--doubled mode--locked Ti:Sapp laser (80 MHz
rep. rate, $\lambda=390$ nm) pumps a 2 mm BBO crystal. Pinholes
$p$ perform spatial selection of the entangled modes. The pump is
reflected onto itself by mirror M3 that is mounted on a
computer--controlled translation stage. Mirrors M1 and M2 form the
feedback loop, including a polarization rotation element
($\lambda/2$), for the entangled photons. Photon detection of the
$\ket{1,1;1,1}$ term in the H/V basis occurs at avalanche photo
diodes D1-D4, after going through polarizing beam splitters (PBS)
and 5--nm--bandwidth filters f1-f4. The role of the two extra
1--mm BBO crystals in modes $a$ and $b$ is to compensate for
undesirable birefringent properties of the main crystal
\cite{kwiat:95}.} \label{fig:easer}
\end{figure}

The set--up used to demonstrate stimulated entanglement is
illustrated in Fig. \ref{fig:easer}. A 120--fs pump pulse at 390
nm (with a repetition rate of 80 MHz) passes through a
$\beta$--barium--borate (BBO) crystal and creates pairs of
polarization--entangled photons in spatially distinct modes $a$
and $b$. The experimental parameters are chosen such that (to
first order) the singlet photon--pair state \ref{eq:spin1singlet}
is created. Initially modes a and b are in the vacuum state and
the photon pairs are spontaneously created. The fact that modes
$a$ and $b$ geometrically diverge, and that horizontally and
vertically polarised photons experience different crystal
parameters, limits the useful crystal length \cite{atature:99} and
thereby prohibits an efficient stimulated emission process. To
obtain significant stimulated emission we redirect the
spontaneously created photon pairs into the crystal at the same
time (tuned by a delay on mirror M3) as the reflected pump pulse
passes through the crystal a second time. Provided that the
feedback loop for the photon pairs is polarization independent,
which is obtained by using a bow-tie folded geometry including a
$\lambda/2$ waveplate that exchanges $H$ and $V$ polarizations,
optimum conditions for stimulated emission of photon pairs can be
established. As stimulated emission can be seen as a constructive
multi--particle interference effect, and because the process of
parametric down--conversion is sensitive to the phase of the pump,
we should expect to observe an oscillation between stimulation and
suppression of emission as function of the pump--pulse delay. The
period of this oscillation corresponds to the optical frequency of
the pump laser. In the region where the difference between the
pump delay and the feedback loop is larger than the coherence
length of the observed photons (determined by the 5--nm
narrow--bandwidth filters in front of the single--photon
detectors) no such interference pattern is expected.

\begin{figure}
\scalebox{.77}{\includegraphics{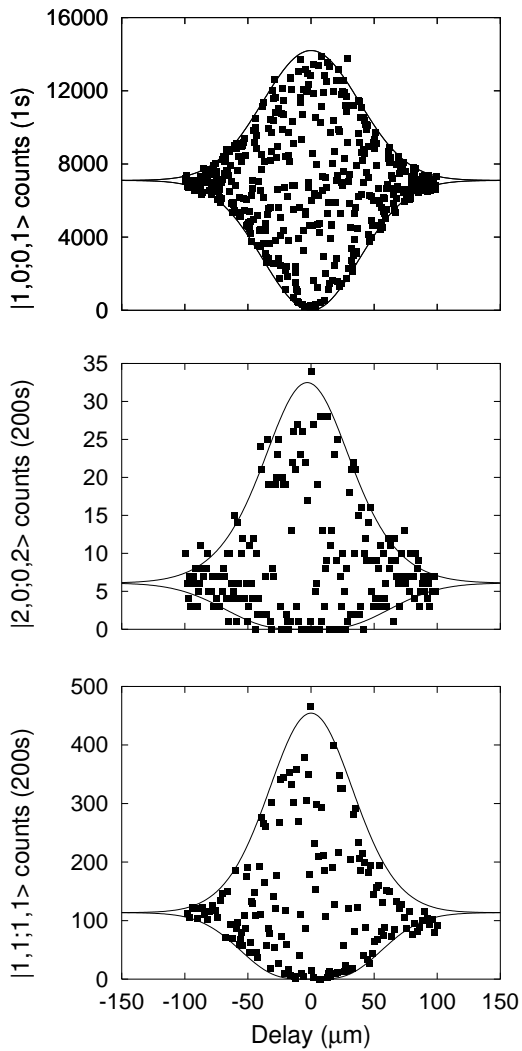}}
     \caption{Experimental demonstration of stimulated entanglement.
     The top graph shows the 2--fold coincidence rate corresponding
     to the detection of the $\ket{1,0;0,1}$  term in the 45°/-45°
     basis as function of the delay between the reflected pump and
     the entangled photons generated in the first pass through the
     crystal. The solid lines are theoretical fits to the envelope
     of the curve as the degree of overlap varies. Similarly the
     middle graph shows the 4--fold coincidence rate corresponding to
     the detection of the $\ket{2,0;0,2}$  term in the 45°/-45° basis
     and the bottom graph the $\ket{1,1;1,1}$  in the H/V basis. The
     effect of stimulated emission is apparent in the increase of the
     number of 4--fold coincidences at zero delay of a factor of 5.3
     and 4.0 for the $\ket{2,0;0,2}$   and $\ket{1,1;1,1}$  terms (see text).
     The difference in rates between the two 4-photon graphs is due
     to the probabilistic detection and extra elements introduced to
     measure the $\ket{2,0;0,2}$  term.} \label{fig:rough}
\end{figure}
\begin{figure}
\scalebox{0.87}{\includegraphics{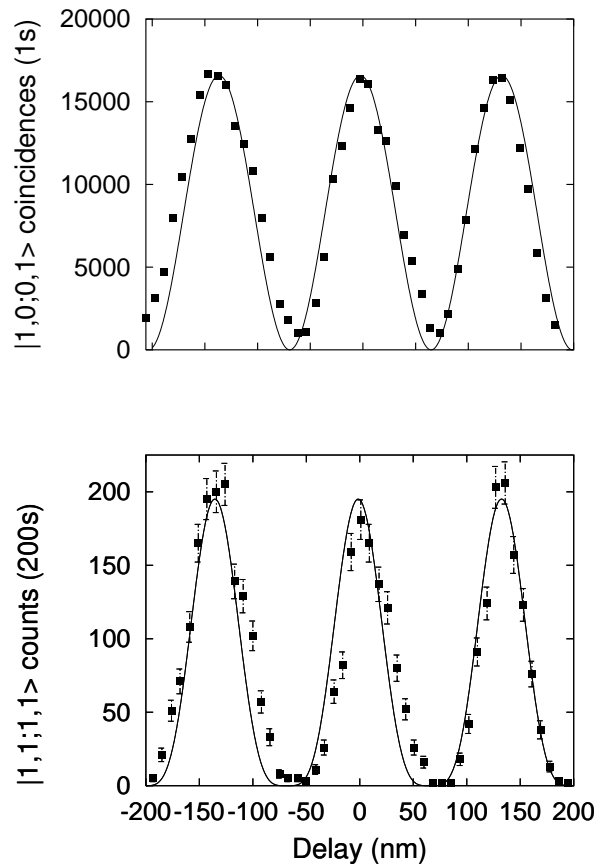}}
     \caption{Two-- and four--photon interference due to stimulated
emission. A fine scan of the  $\ket{1,0;0,1}$ (top) and
$\ket{1,1;1,1}$  (bottom) terms in the zero delay region shows
optimum stimulation and suppression of the 2-- and 4--fold
coincidence probability.} \label{fig:fine}
\end{figure}

To study the 2-- and 4--photon entangled states we measure each
term in equations \ref{eq:singlet} and \ref{eq:spin1singlet}
individually in two non--orthogonal polarization bases. The
$\ket{1,0;0,1}$ and the $\ket{0,1;1,0}$  terms are measured in the
desired bases by using a polarizer in front of a single photon
detector in each of the spatial modes $a$ and $b$. The
$\ket{1,1;1,1}$ term is detected by the introduction of polarizing
beam splitters in the appropriate basis in each mode followed by
four single--photon detectors (see Fig. \ref{fig:easer}). As we do
not use multi--photon detectors, we can only measure the
$\ket{2,0;0,2}$  and the $\ket{2,0;0,2}$ term with a 0.25
probability using a combination of a polarizer, a 50--50 beam
splitter and two single--photon detectors in each mode.

Quantitative predictions for the amplification of the individual
terms in equations \ref{eq:singlet} and \ref{eq:spin1singlet}
resulting from the double pass configuration are obtained by
expanding the unitary evolution of the created light fields in the
polarization--entangled photon-pair creation operator $\Kd$.  To
second order in $\Kd$ we obtain:
\begin{eqnarray}
\hat{U}&=&\hat{U}_2\hat{U}_1=1+e^{i\theta}\tau\Kdt+\tau\Kdo \nonumber \\
&+&\frac{1}{2}e^{2i\theta}\tau^2(\Kdt)^2+\frac{1}{2}\tau^2(\Kdo)^2+e^{i\theta}\tau^2\Kdt\Kdo
\label{eq:expansion}
\end{eqnarray}
where subscripts 1 and 2 refer to the first and second pass
through the crystal. The relative phase $\theta$ between the first
and second pass of the pump pulse is tunable by the translation of
mirror M3 in Fig. \ref{fig:easer}. We discuss two limits: when
$\Kdo=|kdt$ and when $\Kdo$ is distinguishable from $\Kdt$. The
first case applies at zero delay where efficient phase--sensitive
stimulated emission occurs. From equation \ref{eq:expansion} it
follows that doubling the value of the interaction parameter
results in an increase in probability for the 2-photon terms from
$\tau^2$ to $4\tau^2$,  and in an increase for the 4--photon terms
from $\tau^4$ to $16\tau^4$. Note that the 4--photon state has a
four times larger amplification than the 2--photon states, which
is characteristic of stimulated emission. The second case applies
if the reflected pump pulse delay between the two passes is not
equal to the delay of the entangled photons in the feedback loop.
In this case there are simply two independent contributions to the
2--photon detection events, but there are several distinct
contributions to the 4--photon detection events. Each single pass
has a small probability of $\tau^4$  to create state
\ref{eq:spin1singlet}. In addition, as current single photon
detectors do not have a high enough time resolution to distinguish
between photons arriving from the first or second pass, there are
spurious 4--fold coincidences from a combination of 2--photon
states created in both passes. The spurious contributions to the
$\ket{2,0;0,2}$ and $\ket{0,2;2,0}$ the detections will be
$\tau^4$ and for the $\ket{1,1;1,1}$ detection will be $2\tau^4$.

We scan from the region where $\Kdo$  is completely
distinguishable from $\Kdt$ into the region where $\Kdo=\Kdt$,
while observing the intensity of the 2-- and 4--photon terms. From
the considerations above we expect ---in the case of ideal
stimulated emission--- that the terms in Eq. \ref{eq:singlet} to
show a two--fold increase and that the middle term in Eq.
\ref{eq:spin1singlet} to show a four--fold increase, and the other
two terms to increase by a factor of 16/3=5.33. Owing to the
rotational symmetry of states \ref{eq:singlet} and
\ref{eq:spin1singlet}, these predictions are basis independent.

Figure \ref{fig:rough} shows our experimental data for the
detection of the $\ket{1,0;0,1}$ (top) and the $\ket{2,0;0,2}$
(middle) terms measured in the 45° rotated basis, and the
$\ket{1,1;1,1}$ (bottom) term in the H/V basis. The solid curves
are the envelopes of the oscillating functions giving the maximum
and minimum theoretical values for the coincidence rates.
 The experimental data shows an increase of 1.95±0.10 for $\ket{1,0;0,1}$, 5.3±0.6 for
$\ket{2,0;0,2}$ and of 4.1±0.3 for $\ket{1,1;1,1}$. These results
are in good agreement with the predictions discussed above.
Similar results have been obtained in the other bases and for the
$\ket{0,1;1,0}$ and $\ket{0,2;2,0}$ terms, demonstrating the
rotational invariance ---that is, the spin-1/2 and spin-1 singlet
structure--- of states \ref{eq:singlet} and \ref{eq:spin1singlet}.
Additional data indicates an amplification due to the second pass
of 3.95±0.10 for the 2--fold coincidences and of 17±2 for the
4--fold coincidences. This demonstrates the shifting of the
photon--number pair distribution towards terms with higher photon
numbers, a characteristic of stimulated emission. A final proof of
stimulated emission --seen as a constructive interference
process-- is the phase--dependent emission probability shown in
Fig \ref{fig:fine}. This is a fine scan around the region of zero
delay for the $\ket{1,0;0,1}$ (top) and the $\ket{1,1;1,1}$ (bottom)
terms in the 45° rotated basis and H/V basis, respectively. The solid
lines are fits to the theoretical predictions, which vary as
$(1+\cos \theta)$  for the 2--photon case and as $(1+\cos \theta)^2$
for the 4--photon case. The visibility of these interference fringes
is in all cases above 97%.

In summary, we have pointed out the rich entanglement structure
---equivalent to a superposition of spin-$n/2$ singlet states---
obtained by stimulated emission of the familiar rotationally
symmetric Bell state; we have also demonstrated that stimulated
emission of polarization--entangled photons can be achieved
experimentally. Both the characteristic shifting of the
photon--pair distribution towards higher photon numbers, and the
rotational symmetry of the 2-- and 4--photon contribution, have
been observed. The good agreement between the experimental data
and the theory shows that the stringent indistinguishability
requirements to obtain entanglement in a stimulated process using
external resonators have been met. Although related theory and
experiments on interference enhanced emission of photon pairs and
on photon injection into non-linear crystals have been reported
\cite{demartini:00,demartini:01,herzog:94,milonni:96}.Our results
constitute (to our knowledge) the first experimental demonstration
of the onset of laser--like operation for entangled photons.

Using multi--pass amplification pumped by higher--intensity pulses
it should be possible to produce rotationally-symmetric
multi--photon entangled states with an average photon (pair)
number of the order of 100. As exploration of such states is based
on post--selection, the challenge of creating them should go in
parallel with the challenge of constructing low--loss transmission
lines and high--efficiency multi--photon detectors. Although there
are very encouraging developments on low--loss optical fibres
\cite{cregan:99} and highly efficient multi--photon detectors
\cite{kim:99}, we will always have to face the situation of losing
photons in the process of creating, transporting and analyzing the
desired state \ref{eq:hamiltonian}. As this state is one large
complex entangled state, one might think that the loss (or a
measurement) of a single photon will destroy all the interesting
properties of the state. On the contrary, the complex entanglement
has the remarkable feature that the loss or measurement of one (or
more) particles does not eliminate all the entanglement between
the remaining particles. To illustrate this point we focus our
attention on the 4-photon state \ref{eq:spin1singlet} and consider
a measurement of a photon in mode $a$ in the H/V basis, with the
measurement result being H. The state of the remaining three
particles will be $1/\sqrt{2}(\ket{1,0;0,2}-\ket{0,1;1,1})$ which
still contains non--maximal entanglement between modes $a$ and
$b$. A generalization to the actual loss of several photons form
the full state (2) is currently under study.

The increased sharpness of the peaks in the interference pattern
for the singlet of spin--1 (see fig. \ref{fig:fine}) is
potentially of interest for applications in metrology. An
interferometric measurement of relative distance, for example,
would benefit from the more precise location of the maxima
obtained by looking at the 4--fold (or higher) coincidence rates.
Furthermore, the increased amount of entangled terms made
available by stimulated emission and post--selection, offer new
possibilities for higher bit--rate quantum cryptography. We
consider that entanglement robustness ---together with the
rotational symmetry of the state created by stimulated
polarization entanglement--- opens the way to many applications in
quantum information and provides a powerful tool to study the
almost unexplored area between the discreet and the macroscopic
optical quantum correlation experiments.

%\bibliography{fund-tests,PDCexperiments,theorypdc,qcomputation,qcrypto,qinfo,qteleport,qoptics,tecnology}

\end{document}